\documentclass[reprint,superscriptaddress,showkeys,amsmath,amssymb,aps,]{revtex4-2}

\usepackage{graphicx}
\usepackage{dcolumn}
\usepackage{bm}

\newcommand{\ket}[1]{\vert #1\rangle}

\newcommand{\hh}{\hat{H}}
\newcommand{\sz}{\hat{\sigma}_z}

\newcommand{\spp}{\hat{\sigma}_+}
\newcommand{\sm}{\hat{\sigma}_-}

\begin{document}

\preprint{APS/123-QED}

\title{The role of electron polarization on nuclear spin diffusion}

\author{Alessandro Chessari}
\affiliation{%
 Univ. Lyon, CNRS, ENS Lyon, UCBL, Université de Lyon, CRMN UMR 5082, 69100 Villeurbanne, France
}

\author{Samuel F. Cousin}%
\affiliation{Aix Marseille Univ, CNRS, ICR, 13397 Marseille, France}

\author{Sami Jannin}%
\affiliation{%
 Univ. Lyon, CNRS, ENS Lyon, UCBL, Université de Lyon, CRMN UMR 5082, 69100 Villeurbanne, France
}

\author{Quentin Stern}%
 \email{quentin.stern@protonmail.com}
\affiliation{%
 Univ. Lyon, CNRS, ENS Lyon, UCBL, Université de Lyon, CRMN UMR 5082, 69100 Villeurbanne, France
}%

\date{\today}

\begin{abstract}

Dynamic nuclear polarization (DNP) is capable of boosting signals in nuclear magnetic resonance by orders of magnitude by creating out-of-equilibrium nuclear spin polarization. The diffusion of nuclear spin polarization in the vicinity of paramagnetic dopants is a crucial step for DNP and remains yet not well understood. In this Letter, we show that the polarization of the electron spin controls the rate of proton spin diffusion in a DNP sample at 1.2 K and 7 T; at increasingly high electron polarization, spin diffusion vanishes. We rationalize our results using a 2 nucleus - 1 electron model and Lindblad's Master equation, which generalizes preexisting models in the literature and qualitatively accounts for the experimental observed spin diffusion dynamics. 

\end{abstract}

\maketitle

Dynamic nuclear polarization (DNP)\cite{overhauser1953polarization, hall1997polarization, ardenkjaer2003increase,duijvestijn198313c} has enabled numerous applications in nuclear magnetic resonance (NMR) and magnetic resonance imaging (MRI), from real-time imaging of \textit{in vivo} metabolism\cite{golman2006real} to the study of active sites at the surface of catalysts\cite{kobayashi2015dynamic}, to list only two striking examples\cite{plainchont2018dynamic, jannin2019application, wang2019hyperpolarized}. By creating an out-of-equilibrium nuclear polarization, DNP boosts the sensitivity of magnetic resonance by sometimes up to 5 orders of magnitude\cite{ardenkjaer2003increase}, allowing for the detection of phenomena which are otherwise not observable. Many experimental conditions are amenable to solid-state DNP, whether performed on crystals\cite{jeffries1965dynamic,duijvestijn198313c} or frozen liquid\cite{hall1997polarization}, at low\cite{tan2019three} or high magnetic field\cite{rosay2016instrumentation} and all the way down from 1 K\cite{ardenkjaer2003increase} up to room temperature\cite{fujiwara2021triplet}. In all cases, it requires the presence of stable or transient unpaired electron spins\cite{overhauser1953polarization, kundu2019dnp}. The high polarization of the latter is transferred to the surrounding nuclei \textit{via} microwave ($\mathrm{\mu w}$) irradiation close to the resonance frequency of the electron spins. Nuclear spin diffusion then spreads out the polarization from the nuclear spins in the vicinity of the electron spin towards those further away in the bulk\cite{bloembergen1949interaction,khutsishvili1966spin}. However, from the early days of DNP, it was recognized that the nuclei closest to the electron were unable to exchange polarization with their neighbors because their frequency was strongly shifted by the hyperfine interaction (HFI) with the electron because the nuclear dipolar Hamiltonian is truncated by the HFI, which prevents energy conservative nuclear flip-flops\cite{blumberg1960nuclear, khutsishvili1962spin}. These magnetically isolated spins are often said to be within the so-called ‘spin diffusion barrier’. 60 years after this concept has been introduced, it is still not clear at which distance from the electron nuclear spin diffusion starts being effective. Moreover, if diffusion is sometimes shown to be effective even where the dipolar interactions is presumably truncated, the mechanisms behind this phenomenon are not well understood yet\cite{tan2019three, stern2021direct,jain2021dynamic, mr-2022-12}.

\begin{figure}[b]
	\centering
	\includegraphics[width=\columnwidth]{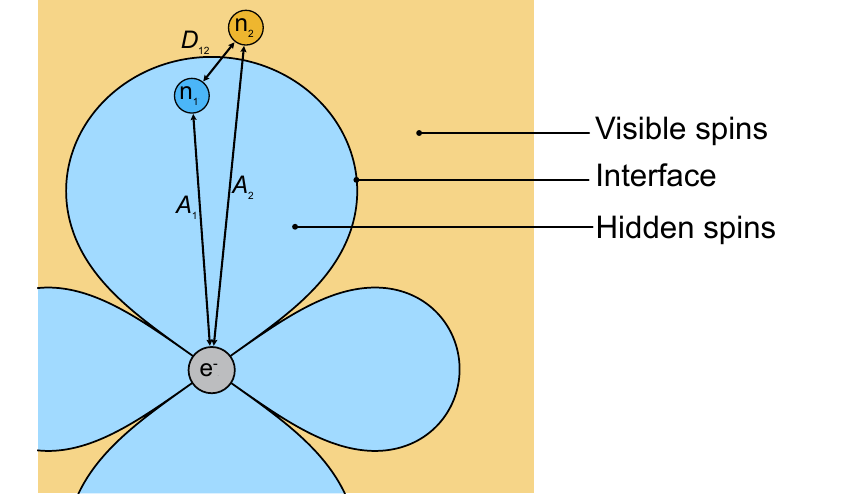}
	\caption{Schematic representation of the hidden and visible nuclear spins interacting with the electron spin through the HFI with intensities \textit{A}$_1$ and \textit{A}$_2$. The two nuclei interact together with the dipole-dipole coupling constant \textit{D}$_{12}$. }	
	\label{Fig_Intro}
\end{figure}

We have recently introduced the hyperpolarization resurgence experiment (HypRes), a method which enables the observation of nuclear spin diffusion from nuclear spins nearby the electron, which are not accessible to direct NMR detection, to those further from the electron, which can be observed by NMR (the hidden and visible spins, respectively), as depicted in Fig. \ref{Fig_Intro}\cite{stern2021direct}. Our measurements with the HypRes method revealed that proton spin diffusion had a strong temperature dependence between 1.2 and 4.2 K. Based on a mechanism proposed by Horvitz\cite{horvitz1971nuclear, wolfe1973direct, buishvili1975nuclear, sabirov1979nuclear, atsarkin1980nuclear}, we postulated that the electron polarization, which varies from $83\%$ at 4.2 K to $99.93\%$ at 1.2 K and 7.05 T could be the key factor explaining this variation in nuclear spin diffusion in our observations\cite{stern2021direct}. 

In this Letter, we confirm our hypothesis both experimentally and theoretically. We introduce an extension of the HypRes experiment where the electron polarization is lowered to a stable level by $\mathrm{\mu w}$ irradiation, which allows comparing the efficiency of spin diffusion for various electron spin polarization. We find that proton spin diffusion at 1.2 K in an amorphous sample doped with TEMPOL is indeed faster when the electron polarization is reduced by $\mathrm{\mu w}$ irradiation. We then rationalize these findings by constructing a model of spin diffusion including one electron spin and two nuclear spins as depicted on Fig. \ref{Fig_Intro}, where the electron is treated as a random fluctuating field\cite{horvitz1971nuclear} and the nuclear spin dynamics are described in terms of Lindblad's master equation\cite{Lindblad1976,gorini1976completely, karabanov2014spin,bengs2020master,pell2021method}. The electron polarization intervenes in this model in two ways. First, because of the fast oscillation of the electron spin state, the nuclear spins experience an averaged HFI, which reduces the frequency mismatch with their neighbor nuclear spins. Second, the oscillation of the electron spin state compensates for the remaining energy mismatch between the $\ket{\alpha\beta}$ and $\ket{\beta\alpha}$ states, thus enabling spin diffusion\cite{horvitz1971nuclear}.

\begin{figure*}[!hbt]
	\centering
	\includegraphics[width=\textwidth]{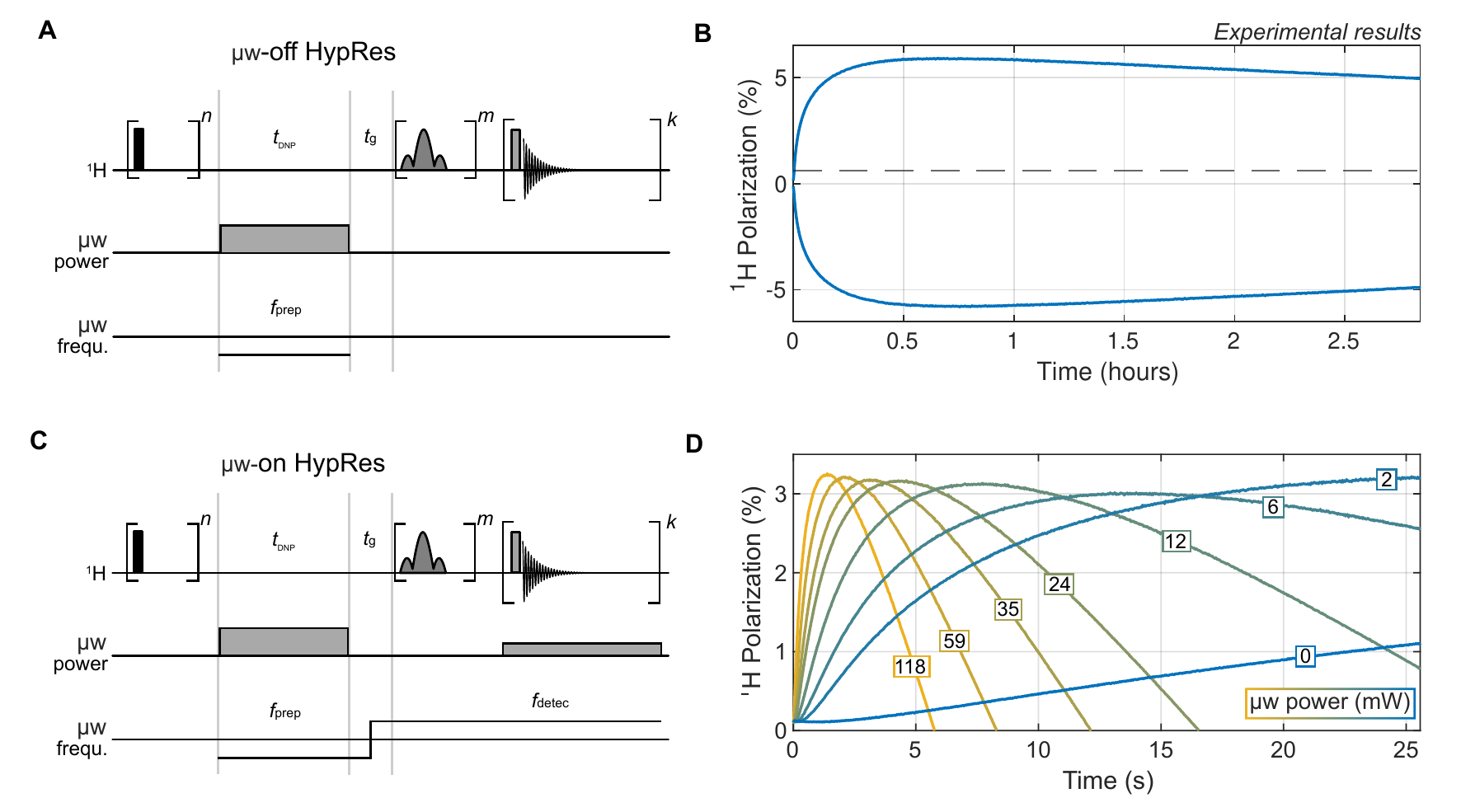}
	\caption{Pulse sequence diagrams (panels A and C, respectively) and results of the $\mathrm{\mu w}$-off and -on HypRes experiments at 7.05 T and 1.2 K (panels B and D, respectively). On the $^1$H channel of the pulse sequence diagram, the black and gray rectangles represent square pulses, with nutation angles of $\pi$/2 and 0.1°, respectively. After an initial strong saturation with $n = 120$ $\pi$/2 pulses, $\mathrm{\mu w}$ irradiation was switched on at the maximum available power during delay $t_{\textrm{DNP}}=$ 640 s of the preparation at a frequency $f_\textrm{prep}$ of 197'648 and 198'112 MHz for positive and negative DNP, respectively, with a saw-tooth frequency modulation of 160 MHz around the central frequency, at a rate of 500 Hz. After switching off $\mathrm{\mu w}$ irradiation, a delay $t_\textrm{g} \approx 10$ s let the electron relax towards Boltzmann equilibrium, before the detection was launched. The train of $m = 19$ sinc pulses separated by random delays saturated the visible spins (see the Supplementary Material for more detail on the saturation scheme). Finally, the HypRes signal was recorded $k = 1024$ times, with 0.1° pulses every 10 s and 0.025 s in the case of the $\mathrm{\mu w}$-off and -on experiments, respectively. The $\mathrm{\mu w}$-off HypRes experiments on panel B were recorded with both positive and negative DNP during preparation, causing a positive and negative polarization overshoot, respectively. The horizontal dashed line on panel B indicates the Boltzmann equilibrium polarization of proton spins. In the case of the $\mathrm{\mu w}$-on HypRes experiments, $\mathrm{\mu w}$ irradiation was switched back on after saturation at a specific power and frequency $f_\textrm{detec}$. The $\mathrm{\mu w}$-on HypRes experiments were recorded with positive DNP during preparation and negative DNP during detection. The $\mathrm{\mu w}$ power applied during detection from 0 to 118 mW is indicated on the curves of panel D. }	
	\label{Fig_Exp}
\end{figure*}

Fig. \ref{Fig_Exp}A shows the pulse sequence of the HypRes experiment, similar to that in our previous work\cite{stern2021direct}.

\begin{enumerate}
    \item	The nuclear polarization is first wiped out by a series of saturation pulses. 
    \item	The nuclear spins are polarized by $\mathrm{\mu w}$ irradiation until they reach DNP equilibrium during delay $t_{\textrm{DNP}}$, setting the $\mathrm{\mu w}$ frequency so as to reach either positive or negative nuclear polarization.
    \item	The $\mathrm{\mu w}$ irradiation is gated and a delay $t_\textrm{g}$ allows the electron spins to return to Boltzmann equilibrium\cite{bornet2016microwave,guarin2021effects}.
    \item	The polarization of the visible spins is wiped out by a selective saturation scheme, optimized so as to saturate over a bandwidth of 200 kHz. The hidden spins are negligibly affected because the bandwidth of the saturation pulses is small compared to their hyperfine shifts. 
    \item	The equilibration of polarization between the visible and hidden spins is monitored using small angle pulses.
\end{enumerate}

We performed the $\mathrm{\mu w}$-off HypRes experiment as in Ref. \cite{stern2021direct} on a sample of 50 mM TEMPOL in H${_2}$O:D${_2}$O:D${_8}$-glycerol 1:3:6 (v/v/v) (DNP juice) at 1.2 K and 7.05 T in a pumped liquid helium bath cryostat\cite{elliott2021practical}. The experiment was performed setting the $\mathrm{\mu w}$ frequency during preparation so as to reach either positive or negative nuclear polarization, with a polarization $P_{\textrm{DNP}}^{\textrm{max}}$ on the order of $+70\%$ and $-70\%$, respectively (see the Supplementary Material for more details on polarization quantification). The positive (or negative) polarization acquired during preparation was wiped out by the saturation pulses only for the visible spins, far from the electron ($<$ $0.2\%$ remaining polarization). The spins closer to the electron retained their invisible polarization, which surged onto the visible spins during the course of detection, causing an observable positive overshoot (or negative, respectively). The two resulting curves are shown in Fig. \ref{Fig_Exp}B. The Boltzmann equilibrium polarization of proton spins is indicated by a dashed line for comparison ($0.60\%$ in these conditions). The experimental traces feature two processes: the equilibration of the hidden and visible spins polarization \textit{via} spin diffusion far beyond Boltzmann equilibrium within $\approx0.5$ h, followed by their slow relaxation towards it. The spin diffusion process monitored during the first part occurs while the electron polarization is that of Boltzmann equilibrium, which is $99.93\%$ in these conditions. 

In order to monitor spin diffusion in the same conditions but with a lower electron polarization, we repeated the experiment using the pulse sequence presented in Fig. \ref{Fig_Exp}C. In this case, after $\mathrm{\mu w}$ irradiation was switched off, the $\mathrm{\mu w}$ frequency was changed from the value yielding positive nuclear spin polarization to that yielding negative polarization. $\mathrm{\mu w}$ irradiation was then switched back on during detection. This experiment was repeated with different values of $\mathrm{\mu w}$ power during detection from 0 to 118 mW of the maximum available power, yielding the curves shown on Fig. \ref{Fig_Exp}D. The $\mathrm{\mu w}$ power used during detection is indicated on the curves. Like for the $\mathrm{\mu w}$-off HypRes experiment, the nuclear polarization acquired by the hidden spins during preparation first surged onto the visible spins causing a positive polarization overshoot. Then, instead of decaying towards thermal equilibrium (only $0.60\%$ polarization), negative DNP started pulling the polarization towards the opposite direction at the negative DNP equilibrium value. Most importantly, the stronger the $\mathrm{\mu w}$ power during the equilibration of polarization between the visible and hidden spins and hence the weaker the electron polarization, the faster the flow of nuclear polarization from the hidden to the visible spins. $\mathrm{\mu w}$ irradiation does not influence the nuclei other than through the electron spins (the sample heating by $\mathrm{\mu w}$ irradiation is less than $10$ mK). Therefore, only the electron dynamics can be responsible for the observed rapid spin diffusion from hidden to visible spins.

In order to understand how electron spin polarization influences nuclear spin diffusion, we calculate the transition rate probability $W$ between the states $\ket{\alpha\beta}$ and $\ket{\beta\alpha}$ of a pair of coupled nuclei, at an internuclear distance $a$, with dipolar coupling constant $D_{12}$, both subject to the dipolar field of an electron spin, at distances $r_i$ and with hyperfine coupling constants $A_{z,i}$, as shown in Fig.~\ref{Fig_Intro}. Following Horvitz\cite{horvitz1971nuclear} the electron is treated semi-classically, taking into account the stochastic time dependence of its state. We also assume the nuclear dipole coupling $D_{12}$ to be weak in front of the broadening of the nuclear levels $\ket{\alpha\beta}$ and $\ket{\beta\alpha}$ due to the interactions with the surrounding nuclei. Since the system is immersed in a strong magnetic field that shifts the spin paired levels $\ket{\alpha\alpha}$ and $\ket{\beta\beta}$ far from that the unpaired subsystem, we can restrict our Hilbert space to the one of the $\ket{\alpha\beta}$ and $\ket{\beta\alpha}$ subspace. The restricted Hamiltonian may be expressed in terms of the Pauli matrices $\spp=\hat{I}^+_{1}\hat{I}^-_{2}$ and $\sz=\hat{I}^z_{1}-\hat{I}^z_{2}$ as follows
\begin{align}
  &\hh_D = -\frac{1}{2}D_{12} (\spp+\sm) \\
  &\hh_{HF}(t) = \frac{1}{2}\Delta(\bar{P}+P'(t)) \sz
\end{align}
where $\Delta=A_{z,1}-A_{z,2}$ is the difference of the HFI. For simplicity, we ignore any diagonal terms giving rise to an overall shift of the two energy levels.
The dynamics of the electron polarization $P(t)$ is decomposed into a static contribution given by the average value $\bar{P}$ and an unbiased signal $P'(t)$, with an autocorrelation function $R(\tau)=(1-\bar{P}^2)e^{-\Gamma_c |\tau|}$, where $1/\Gamma_c$ is the correlation time of the electron spin state\cite{horvitz1971nuclear}.
In the rotating frame given by $U_0=e^{-i\int_0^t \hh'_{HF}(\tau)d\tau} $, the Hamiltonian reads 
\begin{equation}
\hh=\frac{1}{2}\Delta\bar{P} \sz-\frac{1}{2}D_{12}\left(\spp f(t) + \sm f^*(t)\right),
\end{equation}
where $f(t)=e^{i\Delta\int_0^t P'(s)ds}$. Since the electron dynamics are stochastic, the absorption is given by the integral of the spectrum of the modulation $\langle f(t) \rangle_{P'}$~\cite{qubit,book,Gardiner2004QuantumNA}
\begin{equation}
F(\omega)=\frac{2\Gamma_c  (1-\bar{P}^2)\Delta^2}{(\omega^2-(1-\bar{P}^2) \Delta^2)^2+\omega^2 \Gamma_c^2},
\end{equation}
with the spectrum of the energy difference between the nuclear levels 
\begin{equation}
S_{\Gamma_2}(\omega-\Delta\bar{P})=\frac{2\Gamma_2}{(\omega+\Delta\bar{P})^2+\Gamma_2^2}.
\end{equation}
This latter has been calculated taking into account the coupling of a bath of surrounding nuclei, using a Lindblad's Master equation\cite{Lindblad1976,Gorini1976,bengs2020master,pell2021method} (see the Supplementary Material). We find that the transition rate probability is given by the following expression that takes explicitly into account the electron polarization~\cite{main,qubit,noise}
\begin{widetext}
    \begin{equation}
        W=\frac{D_{12}^2}{2} \frac{\Gamma_c(1-\bar{P}^2)\Delta^2+\Gamma_2(\bar{\Gamma}^2+\Delta^2)}{((1-2\bar{P}^2)\Delta^2+\Gamma_2\bar{\Gamma})^2+\Delta^2\bar{P}^2(\Gamma_2+\bar{\Gamma})^2},
        \label{EqMainResult}
    \end{equation}
\end{widetext}
where $\bar{\Gamma}=\Gamma_c+\Gamma_2$ and $\Gamma_2$ is the broadening of the nuclear levels due to the bath of surrounding nuclei. When the electron polarization tends towards unity, the electron flip-flops get frozen out (i. e. $\Gamma_c \rightarrow \infty$ and $\bar{P}\rightarrow1$)\cite{abragam1978principles} and so they stop contributing to the transition rate given by Eq. \ref{EqMainResult}, leading to
\begin{eqnarray}
    W=\frac{D_{12}^2}{2} \frac{\Gamma_2}{\Gamma_2^2+\Delta^2}. 
        \label{EqCommonModel}
\end{eqnarray}
This corresponds to the transition rate between two non-degenerate nuclear levels commonly found in the literature\cite{suter1985spin,ernst1998spin, karabanov2014spin, karabanov2018many}. In the limit of high electron polarization with $\bar{P}\ne 1$ (see the Supplementary Material for more details), Eq. \ref{EqMainResult} simplifies to
\begin{eqnarray}
    W=\frac{D_{12}^2}{2}\frac{1-\bar{P}^2}{\bar{P}^2}\frac{\bar{\Gamma}}{\Delta^2\bar{P}^2+\bar{\Gamma}^2},
\end{eqnarray}
which is equivalent to the transition rate obtained by Horvitz perturbative approach in 1971\cite{horvitz1971nuclear}.

\begin{figure}[t]
	\centering
	\includegraphics[width=\columnwidth]{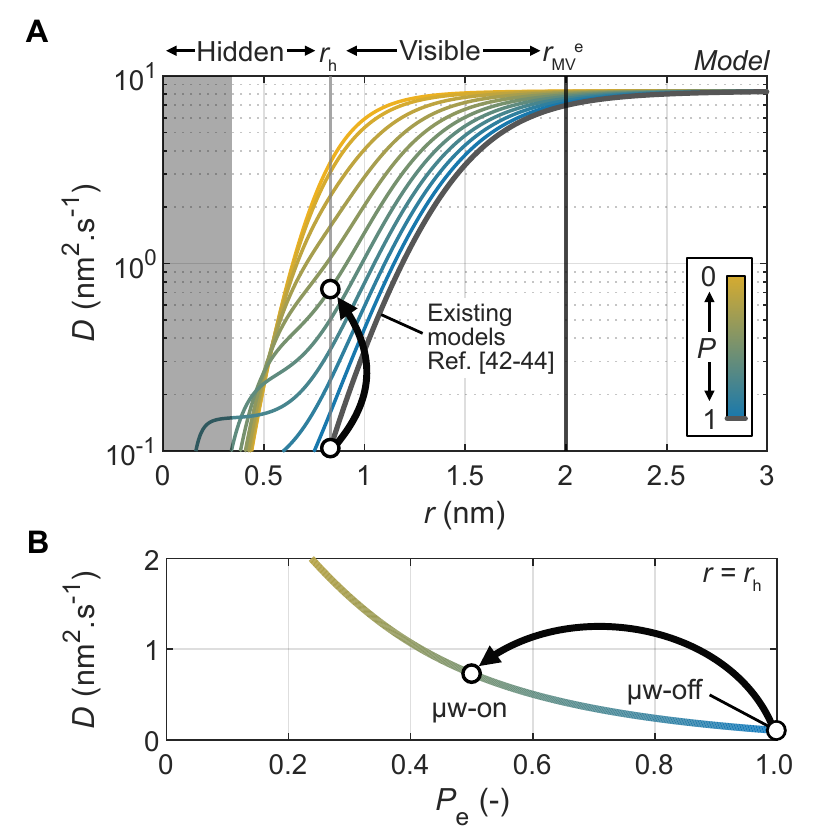}
	\caption{(A) Calculated spin diffusion coefficient as a function of the distance to the electron for the same spin system for different electron polarization $\bar{P}$ between 0 and 1 in steps of 0.1 (see Eq. \ref{EqMainResult}). The vertical black and gray lines represent the radius of the mean volume per electron and the limit between the visible and hidden spins, respectively. The gray area indicates the region where the model hypothesis breaks down ($\Gamma_c < |A_{i,j}|$, see in the text). (B) Spin diffusion coefficient as a function of the electron polarization at the interface between hidden and visible spins. The arrow between the two hollow circles on both panels indicate the increase in diffusion coefficient upon switching on $\mathrm{\mu w}$ irradiation. The nuclear interdistance $a$, the nuclear broadening due to nucleus-nucleus interactions and the electron correlation time $1/\Gamma_c$ are assumed to be 0.66 nm, 18 kHz and 0.5 $\mu$s, respectively.}	
	\label{Fig_Theory}
\end{figure}

Based on the transition rate, we calculate the diffusion coefficient as $D=Wa^2$, with $a$ the average internuclear distance, following Bloembergen\cite{bloembergen1949interaction}. Fig. \ref{Fig_Theory}A shows the diffusion coefficient as a function of the distance of the closest nucleus to the electron for various electron polarizations. The dipolar interaction $D_{12}$ and the HFI $A_{z,1}$ and $A_{z,2}$ were averaged over all orientations so that the diffusion coefficient depends only on $r$ (and not on the angles between the vectors connecting the spins and the magnetic fied). The black and gray vertical lines indicate the radius of mean volume per electron $r_{\textrm{MV}}$ and the radius of the hidden spin reservoir $r_h=0.9$ nm, respectively (which we define as the interface between the hidden and spins, where the proton spins have a coupling of 100 kHz with the electron). The grayed area represents the distance to the electron where the assumption that the nuclear spins experience an average value of the HFI is no longer valid, i. e. $\Gamma_c < |A_{i,j}|$. Fig. \ref{Fig_Theory}B shows the spin diffusion coefficient at the interface between the hidden and visible spins, for an electron polarization from 0 to $100\%$. As depicted by the hollow circles in Fig. \ref{Fig_Theory}, our model predicts that spin diffusion at the interface is $\approx7$ times faster when the electron polarization is $\bar{P} \approx 50\%$ (that is, under $\mathrm{\mu w}$ irradiation\cite{bornet2016microwave,guarin2021effects}) compared to that at Boltzmann equilibrium $\bar{P} = 99.93\%$. As Fig. \ref{Fig_Theory}A shows, the contrast of spin diffuson between $\mathrm{\mu w}$-on and -off is even stronger closer to the electron.

The rate at which polarization rises in the HypRes curves (see  Fig. \ref{Fig_Exp}) is sensitive to the spin diffusion coefficient at the interface between the hidden and visible spins. As we have seen, when the polarization of the electron approaches unity, diffusion is dramatically reduced. In that sense, our theoretical model matches qualitatively with our experimental observations. Because DNP occurs under $\mathrm{\mu w}$ irradiation, that is, at low electron polarization, both our experimental results and our theoretical model lead to the conclusion that spin diffusion in the vicinity of the electron is efficient in our conditions, precisely when DNP is active.

Our model is based on simplifying assumptions which could be improved in several ways. First, we have assumed that the decorrelation rate of the electron spin $\Gamma_c$ is large if front of the HFI but this assumption breaks down at $r = 0.34$ nm (see  Fig. \ref{Fig_Theory}). A more precise calculation would require the use of slow-motion theories\cite{benetis1983nuclear}. Moreover, we have only considered the anisotropic part of the HFI. Improving these points would be necessary if one intends to treat the important case of nuclei on the radical molecule; We have only considered 2-spin order but considering coupled spin terms between more than two spins would lead to predict faster spin diffusion\cite{karabanov2018many}; We have represented the electron spin state using a spectral density function assuming a homogeneous positive electron polarization. Yet, the predicted number of flip-flops could be higher if the non-Zeeman spin temperature of the electron is considered as it leads to significant variation of electron spin polarization along the spectrum of the electron spin, even though with an apparent constant average Zeemann polarization. 

We found that, in this sample, spin diffusion in the vicinity of the electron is effectively quenched when the electron polarization approaches unity. In other samples, other mechanisms which couple the nuclear spins to the lattice phonons could be at play\cite{dolinvsek2000phonon}. In particular, methyl rotation which is still active at temperatures as low as 1 K could contribute to enhancing spin diffusion\cite{meier2013long}. 

In conclusion, we showed both experimentally and theoretically that the rate of proton spin diffusion at 1.2 K and 7.05 T depends on the level of electron polarization: the lower the electron polarization, the faster spin diffusion. To do so, we introduced an extension of the HypRes experiment\cite{stern2021direct} to monitor nuclear spin diffusion in the vicinity of the electron spin while controlling the level of electron polarization \textit{via} $\mathrm{\mu w}$ irradiation. We constructed a model to understand the influence of electron polarization on nuclear spin diffusion. Our model treats both the nuclear relaxation induced by the environment with a Lindblad's master equation approach and the combined effect of a random fluctuating field caused by an electron undergoing flip-flops.

When electron polarization tends towards unity, our model is equivalent to other models where the electron was considered static\cite{karabanov2014spin, karabanov2018many,hovav2012theoretical}. When electron polarization is slightly below unity, in proximity of the electron, our model is equivalent to that of Horvitz \cite{horvitz1971nuclear}. So far, we have only qualitatively compared our model with the experimental results. To further confirm the validity of the proposed model, we intent to model HypRes curves and saturation recovery (either under DNP or at Boltzmann equilibrium) in the future, either combining it with a temperature-like model\cite{prisco2021scaling} or with a full quantum mechanical simulation\cite{karabanov2014spin,karabanov2018many,hovav2012theoretical}. 

Other mechanisms of electron driven nuclear spin diffusion have been reported in the context of MAS-DNP\cite{wittmann2018electron, mentink2017fast, perras2019linear} and $^{13}$C hyperpolarization with colors centers in diamonds\cite{pagliero2020optically}. The mechanism that we have highlighted in this work could potentially be at play in such contexts and therefore contribute to a better understanding of spin diffusion in dDNP as well as in other types of DNP experiments \cite{tan2019three, jain2021dynamic, prisco2021scaling, mr-2022-12}. \\

\begin{acknowledgments}
We thank Christian Bengs for his comments on the manuscript and for insightful discussions on Lindblad's Master Equation. We thank Fr\'ed\'eric Mentink-Vigier and Andrew Pell for enlightening discussions. We acknowledge Bruker Biospin for providing the prototype dDNP polarizer and particularly D. Eshchenko, R. Melzi, M. Rossire, M. Sacher, and J. Kempf for scientific and technical support. We additionally acknowledge C. Jose and C. Pages for use of the ISA Prototype Service, and S. Martinez of the UCBL mechanical workshop for machining parts of the experimental apparatus. This research was supported by ENS-Lyon, the French CNRS, Lyon 1 University, the European Research Council under the European Union’s Horizon 2020 research and innovation program (ERC grant agreement no. 714519/HP4all and Marie Skłodowska-Curie grant agreement no. 766402/ZULF), and the French National Research Agency (project 'HyMag' ANR-18-CE09-0013). 
\end{acknowledgments}

\bibliography{References}

\end{document}


\title{Supplementary Material for\\ ``The role of electron polarization on electron driven spin diffusion''}

\author{{ Alessandro Chessari$^1$, Samuel F. Cousin$^2$, Sami Jannin$^1$ and Quentin Stern$^1$}\\
{\small \em $^1 $Univ. Lyon, CNRS, ENS Lyon, UCBL, Université de Lyon, CRMN UMR 5082, 69100 Villeurbanne, France\\
$^2$Aix Marseille Univ, CNRS, ICR, 13397 Marseille, France
}}

\date{\today}
\maketitle

\section{Polarization quantification}
\begin{figure}[!hbt]
\centering
\begin{subfigure}{.5\textwidth}
    \centering
    \includegraphics[width=.88\linewidth]{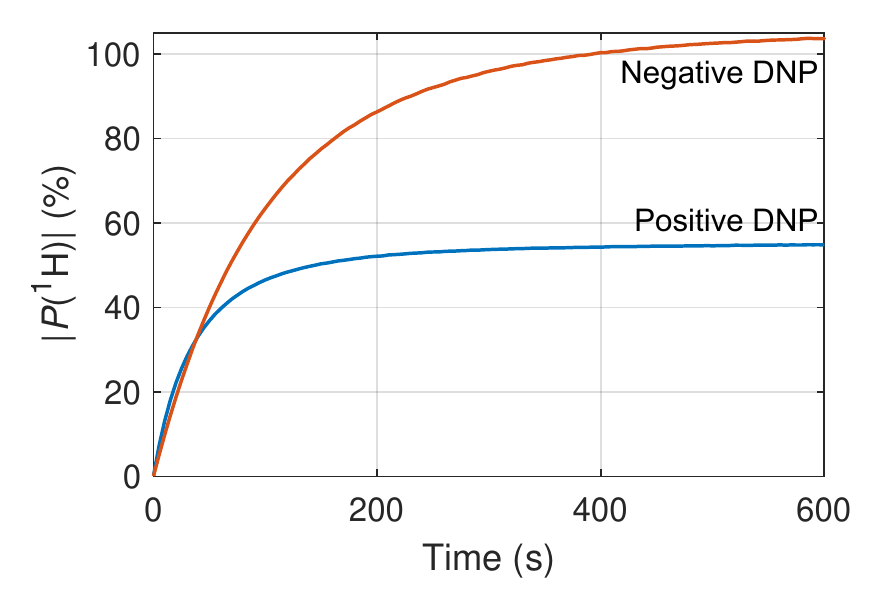}
    \caption{}
    \label{fig:DNPperf_1}
\end{subfigure}%
\begin{subfigure}{.5\textwidth}
    \centering
    \includegraphics[width=.88\linewidth]{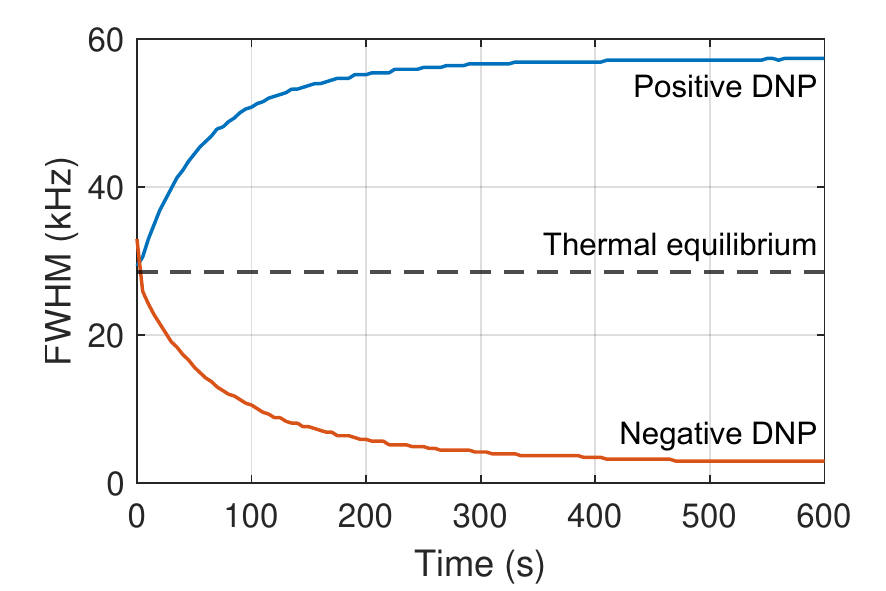}
    \caption{}
    \label{fig:DNPperf_2}
\end{subfigure}
\caption{(\subref{fig:DNPperf_1}) Proton polarization along time under DNP in absolute value both for positive and negative DNP. (\subref{fig:D_vs_Pe_2}) Full width at half maximum (FWMH) of the signal along time for both for positive and negative DNP. The (constant) FWHM of the thermal equilibrium signal is shown for comparison.}
\label{fig:DNPperf}
\end{figure}

~Fig.~\ref{fig:DNPperf_1} shows the proton polarization building up under DNP for positive and negative DNP, monitored using 0.1° pulses. At the same time, the signal broadens and sharpens for positive and negative DNP, respectively, as visible in ~Fig.~\ref{fig:DNPperf_2}. This is the consequence of intense radiation damping which causes a change in apparent nuclear $T_2$ \cite{elliott2021practical}. Because the signal loss during the spectrometer dead time is not negligible, the change in apparent nuclear $T_2$ induces an underestimation and an overestimation of the positive and negative polarization, respectively. The negative polarization is so strongly overestimated that is reaches an nonphysical absolute value above $100\%$. The gap between the positive and negative DNP polarization is mostly an artefact, although a precise quantification is not available in the literature. However, we know from previous works that the steady-state DNP polarization is on the other of $70\%$\cite{stern2021direct}, which corresponds roughly to the average between the negative and positive polarization recorded here.

\section{Determination of the saturation width}
All HypRes experiments in this study were recorded using a saturation scheme consisting of 19 sinc excitation pulses of 100 $\mu s$ with a nutation frequency $\omega_1/2\pi = 51$ kHz, descritized in 1000 points and separated by random delays on the order of $\approx 10$ ms (10.100, 100.105, 100.023, 100.157, 100.101, 100.011, 100.073, 100.054, 100.097, 100.047, 100.026, 100.084, 100.065, 100.105, 100.023, 100.157, 100.101 and 100.011 ms). The sinc pulse shape was generated using Topspin’s built-in functions and chosen to excite over a bandwidth of 200 kHz, with $n = 10$. The profile of the saturation scheme was measured experimentally and confirmed by simulation. 

\begin{figure}[!hbt]
\centering
\begin{subfigure}{.5\textwidth}
    \centering
    \includegraphics[width=.88\linewidth]{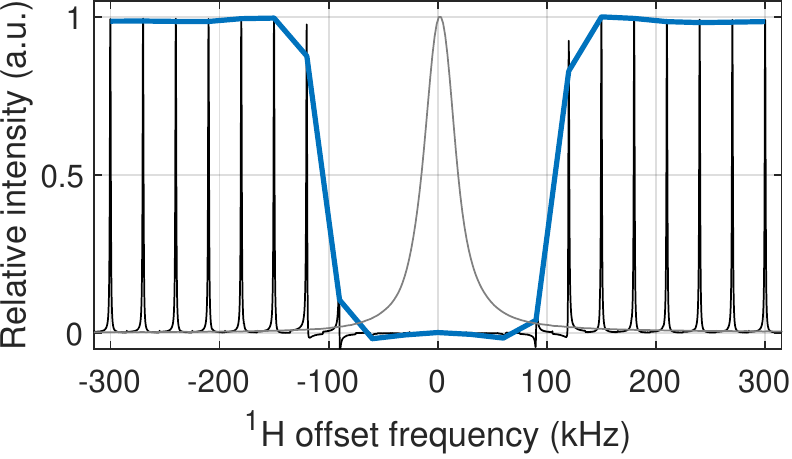}
    \caption{}
    \label{fig:SatOpt_1}
\end{subfigure}%
\begin{subfigure}{.5\textwidth}
    \centering
    \includegraphics[width=.88\linewidth]{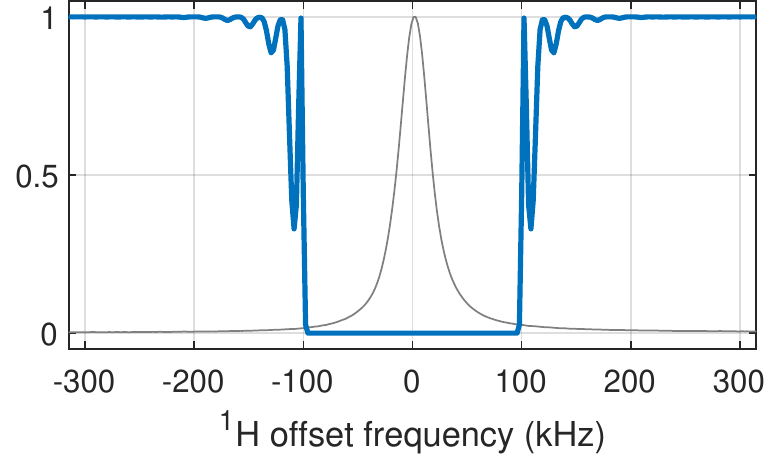}
    \caption{}
    \label{fig:SatOpt_2}
\end{subfigure}
\caption{Measured (\subref{fig:SatOpt_1}) and simulated (\subref{fig:SatOpt_2}) saturation profile of the train of sinc pulses. The faint grey line on both plots represents the NMR signal before saturation. The black signals in (\subref{fig:SatOpt_1}) are the measured signals as a function of the offset of the saturation pulses. The bold blue line in (\subref{fig:SatOpt_1}) is the integral of the measured signals in black. The bold blue line in (\subref{fig:SatOpt_2}) is the simulated remaining polarization along the $z$-axis after saturation.}
\label{fig:SatOpt}
\end{figure}

Experimentally, we first recorded a reference spectrum without saturation (faint gray lines on ~Fig.~\ref{fig:SatOpt}) after 10 s of DNP at 1.2 K. Then we repeated the measurement using the saturation scheme described above varying the carrier frequency of the saturation pulses from -300 to +300 MHz and obtained the traces represented in black on ~Fig.~\ref{fig:SatOpt_1}. The blue line on ~Fig.~\ref{fig:SatOpt_1} is the integral which shows that the saturation width has indeed a bandwidth of 200 kHz.

The excitation profile of the pulse was simulated using a single-spin Hilbert space with the time dependent Hamiltonian
\begin{equation}
    \hh(t) = \Delta \Omega \hat{I}_z +\omega_1 A(t) \{\hat{I}_x\cos{\phi(t)} + \hat{I}_y\sin{\phi(t)}\}
\end{equation}

\noindent where $\Delta \Omega$, $\omega_1$, $A(t)$ and $\phi(t)$ are the offset between the Larmor frequency of the spin and the carrier frequency of the pulse, the nutation frequency of the pulse and the time dependent pulse amplitude (between 0 and 1) and the phase of the pulse generated by Topspin. The initial state of the density matrix was assumed to be $\hat{\rho}_0=\hat{I}_z$. It was propagated under the time dependent Hamiltonian during time steps of 0.1 $\mu$s (corresponding to the discretization of the pulse) using the sandwich formula $\hat{\rho}(t+dt)=\exp{(+i\hh(t)dt)} \hat{\rho}(t) \exp{(-i\hh(t)dt)}$. The remaining polarization along the $z$-axis after the pulse was computed with the trace $P_z=\textrm{Tr}(\hat{I}_z\hat{\rho})/|I|$. The simulation was repeated for 300 offset frequencies $\Delta \Omega$ between -300 and +300 MHz. Plotting the remaining polarization $P_z (\Delta \Omega)$ against the offset frequency $\Delta \Omega$ gives the profile of the remaining polarization along the $z$-axis for a single saturation pulse. To take into account the fact that the saturation scheme consists of 19 pulses, we assumed that the magnetization in the transverse plan decays during the delay between the saturation pulses and so the remaining polarization along the $z$-axis after the $k$th pulse is $P_z ^{final}(\Delta \Omega)=P_z (\Delta \Omega)^k$. The resulting simulated saturation profile is shown on ~Fig.~\ref{fig:SatOpt_2}. It confirms that the excitation scheme saturates the spins from -100 kHz to +100 kHz.

\section{Transition rate}
The Hamiltonian of the system, in the $\ket{\alpha,\beta}$, $\ket{\beta,\alpha}$ subspace, consists of two terms:\\
$\hh(t) = \hh_D+\hh_{HFI}(t)$, where
\begin{align}
\nonumber
  &\hh_D = -\frac{1}{2}D_{12} \sx \\
  &\hh_{HF}(t) = \frac{1}{2}\Delta(\bar{P}+P'(t)) \sz
\end{align}
The first term describes the nucleus-nucleus dipolar coupling, with the coupling given by
\begin{eqnarray}
\nonumber
D_{12}&=&D_0\frac{1}{a_{12}^3}\frac{1-3\cos^2\theta_{12}}{2}\\
D_0&=&\frac{\mu_0}{4\pi}\hbar \gamma_n^2
\end{eqnarray}
where $a_{12}$ the distance between the nuclei and $\cos \theta_{12}$ the angle between the $z$ direction and the vector linking the two nuclei.
The $\hh_{HF}(t)$ term describes the hyperfine coupling between the two nuclei and the electron, which is treated semi-classically~\cite{horvitz1971nuclear}, the hyperfine coupling is given by
\begin{eqnarray}
\nonumber
A_{z,i}&=&A_0\frac{1}{r_i^3}\frac{1-3\cos^2 \theta_i}{2}\\
A_0&=&\frac{\mu_0}{4\pi}\hbar\gamma_n\gamma_e
\end{eqnarray}
where $r_i$ and $\theta_i$ are the distance and the angle between the $i-th$ nucleus and the electron.
We also introduced an additional parameter $\Delta=A_{z,1}-A_{z,2}$ which represents the energy splitting between the nuclear levels in unit of the average electron polarization.
We can describe our system in the interacting picture, given by $\hh_0(t)=\hh_{HF}(t)$ and $\hat{V}=\hh_D$.
In this representation we have $\hh^{(I)}(t)=\hat{U}_0^\dag \hh \hat{U}_0 + i(\partial_t \hat{U}_0^\dag)\hat{U}_0$, with
\begin{eqnarray}
\nonumber
\hat{U}_0(t)&=& \text{exp}\left\{-i\int_0^t d\tau \hh_{0}(\tau)\right\}
\end{eqnarray}
The full Hamiltonian in the interacting picture reduces to
\begin{eqnarray}
\hh^{(I)}(t)=\hat{V}^{(I)}(t)=\hat{U}_0^\dag \hat{V} \hat{U}_0=-\frac{1}{2}D_{12}\sx^{(I)}(t)=-\frac{1}{2}D_{12}[h(\tau)e^{i\Delta\bar{P}\tau}\spp+h^*(\tau)e^{-i\Delta\bar{P}\tau}\sm]
\label{eq:VI}
\end{eqnarray}
with $h(\tau)=e^{i\Delta\int_0^\tau P'(s)ds}$, where we have used
\begin{align}
\hat{\sigma}_\pm^{(I)}(t)&=\hat{U}_0^\dag (t) \hat{\sigma}_{\pm}\hat{U}_0=e^{\pm i\Delta[\bar{P}t + \int_0^t P'(\tau)d\tau]}\hat{\sigma}_{\pm}\\
\sx^{(I)}(\tau)&=\spp^{(I)}(t)+\sm^{(I)}(t)
\label{eq:sigma_interacting}
\end{align}
From the Schrödinger equation, we get for the time evolution operator, for the first-order in the perturbation parameter $D_{12}$
\begin{eqnarray}
\hat{U}^{(I)}(t)&=&1-i\int_0^t d\tau \hat{V}^{(I)}(\tau) + \mathcal{O}(D_{12}^2)
\end{eqnarray}
To obtain the transition rate probability between the state $\ket{\alpha,\beta}$ and $\ket{\beta,\alpha}$ we study the evolution of the projection on $\ket{\beta,\alpha}$ of the evolution of a state prepared in $\ket{\alpha,\beta}$, i. e.
\begin{eqnarray}
c_{\ket{\alpha,\beta} \rightarrow \ket{\beta,\alpha}}(t)= \braket{\beta,\alpha|\hat{U}(t)|\alpha,\beta}=\tensor[_I]{\braket{\beta,\alpha|\hat{U}^{(I)}(t)|\alpha,\beta}}{_I}
\end{eqnarray}
Since the perturbation is off diagonal, we can sum over all the final state and, making use of the completeness relation $\sum_{\ket{m}=\ket{\alpha,\beta},\ket{\beta,\alpha}}\ket{m}\bra{m}=1$, write for the transition probability
\begin{align}
P_{\ket{\alpha,\beta} \rightarrow \ket{\beta,\alpha}}&=|c_{\ket{\alpha,\beta} \rightarrow \ket{\beta,\alpha}}(t)|^2=\sum\limits_{\ket{m}=\ket{\alpha,\beta},\ket{\beta,\alpha}}|c_{\ket{\alpha,\beta} \rightarrow \ket{m}}|^2= \sum\limits_{\ket{m}=\ket{\alpha,\beta},\ket{\beta,\alpha}} \tensor[_I]{\braket{\alpha,\beta|\hat{U}^{(I)}(t)|m}}{_I} \tensor[_I]{\braket{m|\hat{U}^{(I)}(t)|\alpha,\beta}}{_I}\\
&=\text{Tr}\{ \rho_0^{(I)} \hat{U}^{(I)}(t) \hat{U}^{(I)}(t) \} =\frac{D_{12}^2}{4}\int_0^t\int_0^t dt' dt'' \langle \sx (t')  \sx (t'') \rangle
+ \mathcal{O}(D_{12}^3)
\end{align}
where the last average is calculated with respect the density matrix $\hat{\rho}_0=\ket{\alpha,\beta}\bra{\alpha,\beta}$.
We can now perform the variable change $s=\frac{1}{2}(t'+t'')$, $\tau=t''-t'$ 
\begin{eqnarray}
P_{\ket{\alpha,\beta} \rightarrow \ket{\beta,\alpha}}=\frac{D_{12}^2}{4}\int_0^t ds \int_{-B(s)}^{B(s)} d\tau \langle \sx(s+\tau/2) \sx (s-\tau/2) \rangle
\label{eq;partial}
\end{eqnarray}
with
\begin{equation}
B(s)=  
\begin{cases*}
s-t & if $s\leqslant\frac{t}{2}$,\\
s & if $s>\frac{t}{2}$
\end{cases*}
\end{equation}
Now, since the time evolution operator contains a random process, the average in Eq.\eqref{eq;partial} has to be intended also as an average over all the realization of the process $P'(t)$. With this in mind, we can exploit the time translation symmetry of the Hamiltonian and, in the limit of large $t$, the integration gives
\begin{eqnarray}
W_{\ket{\alpha,\beta} \rightarrow \ket{\beta,\alpha}}=\frac{d}{dt} P_{\ket{\alpha,\beta} \rightarrow \ket{\beta,\alpha}}=\frac{D_{12}^2}{4} \int_{-\infty}^{\infty} d\tau \langle \sx (\tau) \sx (0) \rangle_{P'}
\end{eqnarray}
The result is further simplified by the algebra of the Pauli matrices, 
\begin{eqnarray}
W_{\ket{\alpha,\beta} \rightarrow \ket{\beta,\alpha}}=\frac{D_{12}^2}{4} \int_{-\infty}^{\infty} d\tau \langle \hat{\sigma}_- (\tau) \hat{\sigma}_+ (0) \rangle_{P'}
\label{eq:transition_probability}
\end{eqnarray}
To calculate this correlator, we need to break down the calculation into two steps that will be addressed in the two following sections:
\begin{itemize}
    \item First, we need to average over the density matrix which did not take into account until now the effects of the environment. In particular, the density matrix is $\rho_0$ only at $t \rightarrow -\infty$, and its actual time evolution can be obtained tracing out the environmental degrees of freedom. To do this, we will assume a particular realization of the process $P'(t)$ in the following section. The key part of this section will be the a Lindblad Master equation and the quantum regression theorem
    \item Then, we need to do also the averaging over all the processes taking into account the stochastic behaviour of the electronic polarization, i.e. the electronic flip-flop
\end{itemize}

\section{Nuclear Spin Relaxation}
Following~\cite{superoperator} we will first describe generally how, for a system coupled to a bath, the environment degrees of freedom can be traced out. 
Our goal here is to describe nuclear relaxation for a pair of nuclear spins in a bath of surrounding nuclei coupled to the system via the dipolar coupling. We will not consider any relaxation mechanism leading our subsystem to a paired nuclear spin state since these states are split away in energy due to the Zeeman interacton.
To do so, the nuclear bath will be described with a Jaynes-Cummings model, a well-known model in quantum optics~\cite{book}. We will show that within this simple model we can derive the Lindblad operator and the Lindblad Master equation.
Since we are considering a reduced system with a basis composed of only unpaired nuclear spins, our result will be slightly different from the form in which the Lindblad Master Equation as found in the NMR literature~\cite{bengs2020master,karabanov2014spin,karabanov2018many} since the Lindblad operator will be now applied to $\spp=\hat{I}^+_1\hat{I}^-_2$ and $\sm=\hat{I}^+_2\hat{I}^-_1$ rather than on the single spin raising and lowering operators $\hat{I}^+_k$ and $\hat{I}^-_k$.
\\
Let us assume that our total system lives in the Hilbert space $\mathcal{H}_T$ divided into the
environment space living in $\mathcal{H}_E$ and our system living in $\mathcal{H}$. The Liouville-Von Neumann equation gives us the evolution of the density matrix
\begin{eqnarray}
\dot{\hat{\rho}}_T(t)=-i[\hat{H}_T,\hat{\rho}_T(t)]
\end{eqnarray}
where the total system Hamiltonian $\hat{H}_T$ can be written as $\hat{H}_T=\hat{H} \otimes \mathds{1}_E+\mathds{1} \otimes \hat{H}_E +\alpha \hat{H}_{SE}$, being $\hat{H}_E \in \mathcal{L}[\mathcal{H}_E]$ the Hamiltonian of the environment and $\hat{H}_{SE} \in \mathcal{L} [\mathcal{H}_T]$ the Hamiltonian of the system-environment interaction, which is characterised by the constant $\alpha$. We can define the interacting picture by $H+H_{E}$, such that the time evolution of the total density matrix is given by
\begin{eqnarray}
\frac{d}{dt} \hat{\rho}^{(I)}_T(t)&=&-i\alpha[\hh_{SE}^{(I)}(t),\hat{\rho}^{(I)}_T(t)]
\label{eq:evolution_density_matrix}
\end{eqnarray}
This equation can be formally solved to get
\begin{eqnarray}
\hat{\rho}_T^{(I)}(t)&=&\hat{\rho}_T^{(I)}(0)-i\alpha\int_{0}^t d\tau [\hh_{SE}^{(I)}(\tau),\hat{\rho}^{(I)}_T(\tau)]
\label{eq:first_sol_density_matrix}
\end{eqnarray}
Then, inserting Eq.~\eqref{eq:evolution_density_matrix} into Eq.~\eqref{eq:first_sol_density_matrix}
\begin{eqnarray}
\frac{d\hat{\rho}^{(I)}_T(t)}{dt}&=&-i\alpha[\hh^{(I)}_{SE}(t),\hat{\rho}^{(I)}_T(0)]-\alpha^2\int_0^t d\tau [\hh_{SE}^{(I)}(t),[\hh_{SE}^{(I)}(\tau),\hat{\rho}_T^{(I)}(\tau)]]
\end{eqnarray}
Inserting in the latter again the zero-th order expansion for $\rho^{(I)}(\tau)$ around $t$ we get
\begin{eqnarray}
\frac{d\hat{\rho}^{(I)}_T(t)}{dt}&=&-i\alpha[\hh^{(I)}_{SE}(t),\hat{\rho}^{(I)}_T(0)]-\alpha^2\int_0^t d\tau [\hh^{(I)}_{SE}(t),[\hh_{SE}^{(I)}(\tau),\hat{\rho}^{(I)}_T(t)]]+\mathcal{O}(\alpha^3)
\label{eq:evolution_density_matrix_tot}
\end{eqnarray}
Finally, ignoring high-orders terms, the evolution of the density matrix of the system may be obtained tracing out the environmental degrees of freedom
\begin{eqnarray}
\frac{d\hat{\rho}^{(I)}(t)}{dt}&=&-i\alpha \text{Tr}_E[\hh_{SE}^{(I)}(t),\hat{\rho}^{(I)}_T(0)]-\alpha^2\int_0^t d\tau \text{Tr}_E[\hh_{SE}^{(I)}(t),[\hh_{SE}^{(I)}(\tau),\hat{\rho}^{(I)}_T(t)]]
\end{eqnarray}
Now, the interacting system-environment Hamiltonian can be written, without loss of generalities as
\begin{eqnarray}
\hh_{SE}=\sum\limits_i S_i \otimes E_i
\end{eqnarray}
where $S_i$ and $E_i$ are operators acting on the system and on the environment, respectively.
Given this decomposition, Eq.~\eqref{eq:evolution_density_matrix} can be always be rewritten as~\cite{superoperator}
\begin{eqnarray}
\frac{d\hat{\rho}^{(I)}(t)}{dt}&=&-\alpha^2\int_0^t d\tau \text{Tr}_E[\hh_{SE}^{(I)}(t),[\hh_{SE}^{(I)}(\tau),\hat{\rho}^{(I)}_T(t)]]
\end{eqnarray}
We may also suppose that the system and the environment are noncorrelated during all the time evolution, i.e. that the correlation and relaxation time scale are much smaller than that of the system. In this regime we can assume the environment state to be stationary and decoupled from the system state, then $\hat{\rho}_E^{(I)}=\hat{\rho}_E(0)$ and $\hat{\rho}_T^{(I)}=\hat{\rho}^{(I)}(t) \otimes \hat{\rho}_E(0)$. Extending the upper integration limit to infinity and by the change of variable $\tau \rightarrow t-\tau$ leads to
\begin{eqnarray}
\frac{d\hat{\rho}^{(I)}(t)}{dt}&=&-\alpha^2\int_0^\infty d\tau \text{Tr}_E[\hh_{SE}^{(I)}(t),[\hh_{SE}^{(I)}(t-\tau),\hat{\rho}^{(I)}(t) \otimes \hat{\rho}_E(0)]]
\label{eq:ridberg}
\end{eqnarray}
At this point we need to choose the environment Hamiltonian and the coupling term between the system and the bath for our specific case.
The simplest idea is to consider all the surrounding nuclei as a set of quantum harmonic oscillators, following the so-called Jaynes-Cummings model. The bath Hamiltonian is then  $\hh_E=\sum\limits_j \nu_j \hat{a}^\dag_j \hat{a}_j$~\cite{main}, where $\nu_j$ are the frequency and $\hat{a}^\dag_j$ the annihilation operator for the $j$th-oscillator.
We assume that the environment induces transitions in the system by coupling the raising and lowering operators $\spp$ and $\sm$. On the environment the corresponding effect is the annihilation and creation of an energy mode of the oscillators.
The system-environment interaction may then be written as follows
\begin{eqnarray}
\hh_{SE}=\sum\limits_j g_j (\hat{a}^\dag_j \hat{\sigma}_-+\hat{a}_j \hat{\sigma}_+)=\hf\hq ^\dag + h. c.
\label{eq:coupling_environment}
\end{eqnarray}
where $g_j$ represents the coupling strength between the oscillators and the system. The operator $F$ and $Q$ are defined as follows
\begin{equation}
 \nonumber
  \hf = \sm \quad  \hq = \sum\limits_k g_k\hat{a} 
\end{equation}
That, in the interacting picture of $H+H_E$, writes
\begin{equation}
\hf^{(I)}(t)=e^{-i\omega_* t} \hf =e^{- i\Delta(\bar{P}t + \int_0^t P'(\tau)d\tau)}\sm \qquad \qquad
\hq^{(I)}(t)=\sum\limits_k g_k\hat{a}_k(t) =\sum\limits_k g_k\hat{a}_k e^{-i\nu_k t}
\end{equation}
where, to make the notation easier, we have defined  $\omega_*= \Delta \bar{P}+\frac{1}{t}\int_0^t P'(\tau)d\tau $.\\
Taking into account the following relations, where $n_B$ is the Bose-Einstein distribution
\begin{align}
\nonumber
&\text{Tr}_E \left\{ \hrho_E \hat{a}_k(t)\hat{a}_{k'}(t-\tau) \right\}=0\\
&\text{Tr}_E \left\{ \hrho_E \hat{a}_k^\dag(t)\hat{a}_{k'}(t-\tau) \right\}=\delta_{kk'} \langle \hat{a}^\dag_k \hat{a}_k \rangle e^{+i\nu_k t}e^{-i\nu_{k'}(t-\tau)}= \delta_{kk'}\langle \hat{a}^\dag_k \hat{a}_k \rangle e^{+i\nu_k\tau}=\delta_{kk'} n_B(\nu_k) e^{i\nu_k\tau} \\
&\text{Tr}_E \left\{ \hrho_E \hat{a}_k(t)a_{k'}^\dag(t-\tau) \right\}=\delta_{kk'} \langle \hat{a}_k \hat{a}^\dag_k \rangle e^{-i\nu_k t}e^{+i\nu_{k'}(t-\tau)}=\delta_{kk'}\langle \hat{a}_k \hat{a}^\dag_k \rangle  e^{-i\nu_k\tau}=\delta_{kk'}(1+n_B(\nu_k))e^{-i\nu_k\tau}\\
&\text{Tr}_E \left\{ \hrho_E \hat{a}^\dag_k(t)a_{k'}^\dag(t-\tau) \right\}=0
\end{align}
Eq.~\eqref{eq:ridberg} becomes (here we avoid the superscript indicating the interacting picture for simplicity)
\begin{dmath}
\frac{d\hrho(t)}{dt}=\int_0^\infty d\tau\{ 
\hf^{\dag}(t-\tau)\rho(t)\hf(t)\text{Tr}_E [\hq(t-\tau)\rho_E \hq^{\dag} (t)] - \hf(t)\hf^\dag (t-\tau)\rho(t) \text{Tr}_E [\hq^\dag (t) \hq(t-\tau)\rho_E] + \hf(t-\tau)\rho(t)\hf^\dag (t) \text{Tr}_E
[\hq^\dag (t-\tau)\rho_E \hq(t)] - \hf^\dag(t)\hf(t-\tau)\rho(t) \text{Tr}_E[\hq(t)\hq^\dag (t-\tau)\rho_E] +h.c
\}
\end{dmath}
that further simplify to
\begin{dmath}
\frac{d\hrho(t)}{dt}=\int_0^\infty d\tau\{ 
[\hf^{\dag}(t-\tau)\rho(t)\hf(t) - \hf(t)\hf^\dag (t-\tau)\rho(t) ] \langle \hq^\dag (t) \hq(t-\tau) \rangle_E + [\hf(t-\tau)\rho(t)\hf^\dag (t) - \hf^\dag(t)\hf(t-\tau)\rho(t) ] \langle \hq(t)\hq^\dag (t-\tau) \rangle_E +h.c
\}
\end{dmath}
and finally to
\begin{align}
\frac{d\hrho(t)}{dt}=
&[2\hf^{\dag}(t)\rho(t)\hf(t) - {\hf(t)\hf^\dag (t),\rho(t)} ] \int_0^\infty d\tau  e^{-i\omega_* \tau}\langle \hq^\dag (t) \hq(t-\tau) \rangle_E\\ &+[2\hf(t)\rho(t)\hf^\dag (t) - {\hf^\dag(t)\hf(t),\rho(t)} ] \int_0^\infty d\tau e^{i\omega_* \tau} \langle \hq(t)\hq^\dag (t-\tau) \rangle_E
\end{align}
Now, introducing the rate constants
\begin{align}
\Gamma_+(\omega_*)&=4\int_0^\infty d\tau  e^{-i\omega_* \tau}\langle \hq^\dag (t) \hq(t-\tau) \rangle_E=4\pi g(\omega_*)^2_{*}\rho_{D}(\omega_*)n_B(\omega_*) \\
\Gamma_-(\omega_*)&=4\int_0^\infty d\tau  e^{i\omega_* \tau}\langle \hq (t) \hq^\dag(t-\tau) \rangle_E=4\pi g(\omega_*)^2_{*}\rho_{D}(\omega_*)(1+n_B(\omega_*))
\end{align}
where we have considered the levels of the bath as a continuum of states with a density of states give by $\rho_{D}$.
The result, back in the Schrödinger picture is
\begin{eqnarray}
\frac{d\hrho}{dt}&=&-i[\hat{H}(t),\hrho(t)]+\frac{\Gamma^*_+}{2} \mathcal{D}(\spp)\hrho(t)+\frac{\Gamma^*_-}{2} \mathcal{D}(\sm)\hrho(t)
\end{eqnarray}
where we have defined $\Gamma^*_+=\Gamma_+(\omega_*)$, $\Gamma^*_-=\Gamma_-(\omega_*)$ and the dissipator
\begin{eqnarray}
\mathcal{D}(\hat{X})\hrho&=& \hat{X}\hrho\hat{X}^\dag-\frac{1}{2}\{\hat{X}^\dag\hat{X},\hrho\}
\end{eqnarray} 
With a similar procedure, considering a coupling with $\sz$ in Eq.~\eqref{eq:coupling_environment} (we will not enter the details of the dephasing term, we will suppose that mechanisms that can lead to such a coupling exist) we can finally get the Master equation in the Lindblad form $\partial_t \rho(t)=\mathcal{L}\rho(t)$ where
\begin{eqnarray}
\mathcal{L}\rho(t)&=&-i[\hat{H}(t),\hrho(t)]+\frac{\Gamma^*_+}{2} \mathcal{D}(\spp)\hrho(t)+\frac{\Gamma^*_-}{2} \mathcal{D}(\sm)\hrho(t)+\Gamma_2 \mathcal{D}(\sz)\hrho(t)
\label{eq:master_eq}
\end{eqnarray}
This Master Equation is the same as the Lindblad Master equation as found in recent NMR literature~\cite{bengs2020master,karabanov2014spin,karabanov2018many}, reduced in the $\ket{\alpha,\beta},\ket{\beta,\alpha}$ subspace.  
Our simple model used to derive it has all the essential ingredients to describe relaxation: the nuclear relaxation constants $\Gamma^*_+$, $\Gamma^*_-$ and $\Gamma_2$. Any other relaxation mechanisms can be taken into account in the same way we did so far and will lead to corrections to the relaxation rates.\\
Now, we can exploit the quantum regression theorem~\cite{main}\cite{book} that states that, if the equations of motions for the expectation values of a set of system operators are expressed as
\begin{eqnarray}
\frac{d}{dt} \langle \hat{A}_i(t) \rangle=\sum\limits_j G_{ij}\langle \hat{A}_j(t) \rangle
\end{eqnarray}
then for the correlations hold
\begin{eqnarray}
\frac{d}{d\tau} \langle \hat{A}_i(t+\tau)\hat{A}_k(t) \rangle=\sum\limits_j G_{ij} \langle \hat{A}_j(t+\tau)\hat{A}_k(t) \rangle
\label{eq:quantum_regression}
\end{eqnarray}
In the weak coupling regime $D_{12} \ll \Gamma_2$~\cite{milburn}, the evolution of $\langle \sm(t) \rangle$ is given, through the Lindblad Master equation presented in Eq.~\eqref{eq:master_eq} $\partial_t \rho^{(I)}(t)= \mathcal{L}^{(I)} \rho^{(I)}(t)$ (we put ourself in the interaction picture given by $H_0$ as in the last section for simplicity)
\begin{align}
\frac{d}{dt} \langle \sm(t) \rangle&=\text{Tr} \left\{ \frac{d\hrho^{(I)}(t)}{dt} \sm^{(I)}(t) \right\} \\
&=
\text{Tr}\left\{
\left[
-i[\hat{V}^{(I)}(t),\hrho^{(I)}(t)]+\frac{\Gamma^*_+}{2} \mathcal{D}(\spp^{(I)})\hrho^{(I)}(t)+\frac{\Gamma^*_-}{2} \mathcal{D}(\sm^{(I)})\hrho^{(I)}(t)+\Gamma_2 \mathcal{D}(\sz^{(I)})\hrho^{(I)}(t)
\right]
\sm^{(I)}(t)
\right\} \\
&=-\left( \frac{\Gamma^*_+}{2}+\frac{\Gamma^*_-}{2}+\Gamma_2 \right) \langle \sm(t) \rangle \\
\end{align}
By means of~\eqref{eq:sigma_interacting} and of~\eqref{eq:quantum_regression}, neglecting $\Gamma_+^*$ and $\Gamma_-^*$ in front of $\Gamma_2$ (i.e., assuming that nuclear spin-spin relaxation is fast in front of nuclear spin-lattice relaxation)
\begin{eqnarray}
\langle \sm(\tau)\spp(0) \rangle&=&e^{ - \Gamma_2 |\tau| }e^{-i\Delta \bar{P}\tau-i\Delta\int_0^\tau P'(s)ds}
\label{eq:sigma_correlation1}
\end{eqnarray}
where we have set the initial value $\langle \sm(0)\spp(0) \rangle=1$, that corresponds to our choice for the density matrix $\hat{\rho}_0=\ket{\alpha,\beta}\bra{\alpha,\beta}$. Similarly, we obtain that all the others correlations are null.

\section{Electronic flip-flop}
\begin{figure}[!hbt]
\centering
\begin{subfigure}{.5\textwidth}
    \centering
    \includegraphics[width=.88\linewidth]{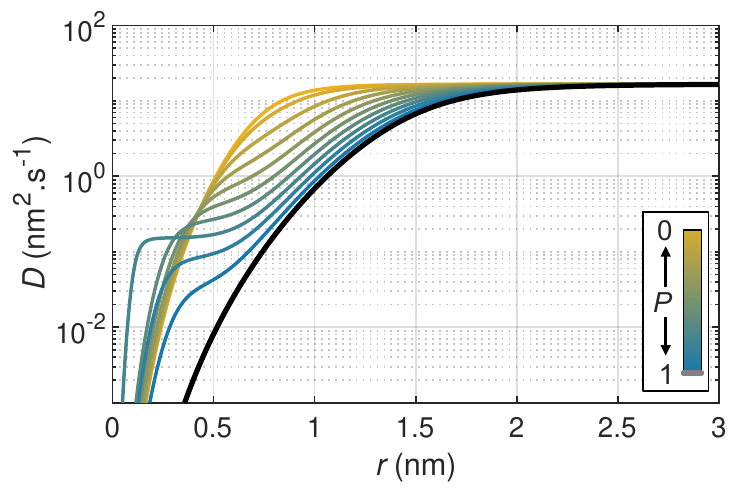}
    \caption{}
    \label{fig:D_vs_Pe_1}
\end{subfigure}%
\begin{subfigure}{.5\textwidth}
    \centering
    \includegraphics[width=.88\linewidth]{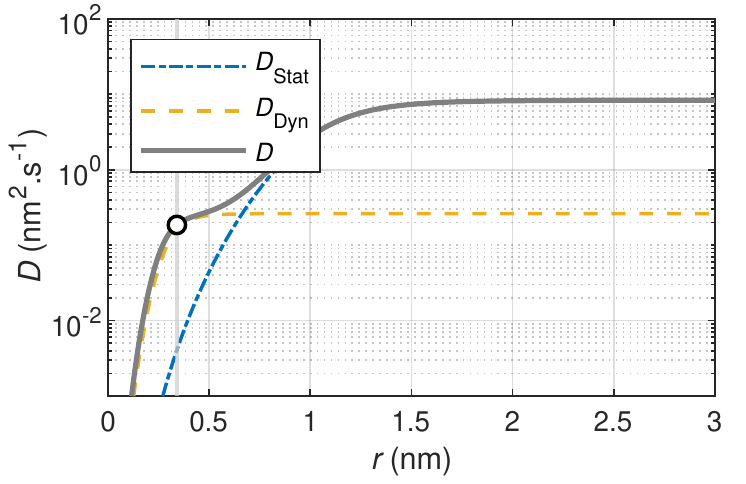}
    \caption{}
    \label{fig:D_vs_Pe_2}
\end{subfigure}
\caption{In~(\subref{fig:D_vs_Pe_1}) the diffusion coefficient in function of the distance from the electron at different electron polarization. In~(\subref{fig:D_vs_Pe_2}) the diffusion coefficient and its limit cases at $\bar{P}=0.6$, where we have marked the diffusion coefficient at $r=\SI{0.34}{nm}$.
Both plots are obtained with the following parameters: $\Gamma_2=\SI{9}{kHz}$, $\Gamma_c=\SI{2}{MHz}$, $a_{12}=\SI{0.66}{nm}$.}
\label{fig:D_vs_Pe}
\end{figure}

Finally, we will consider the electron motion not as a deterministic signal, but rather as undergoing stochastic \textit{flip-flop}, i.e. instantaneously changing in its polarisation, along the $z$ axis with average rate $\Gamma_c$. If $T_{1e}$ and $T_{2e}$ are the spin-lattice and spin-spin relaxation time of the electron, respectively, the $\Gamma_c$ in an insulating solid is \cite{pell2019paramagnetic}
\begin{equation}
\Gamma_c=\frac{1}{T_{1e}}+\frac{1}{T_{2e}}
\end{equation}
In our conditions, the $1/T_{2e}$ is large in front of $1/T_{1e}$ due to strong electron-electron interactions and so we have $\Gamma_c=1/T_{2e}$. However, the derivation presented here holds regardless of the origin of the fluctuations of the electron spin state. Being $\bar{P}$ the average electron polarization, we can write
\begin{equation}
P(t)=P'(t)+\bar{P}
\end{equation}
being $P'(t)$ the unbiased process.
From~\cite{horvitz1971nuclear} we have that the power spectral density of the unbiased process is
\begin{equation}
S_{P'}(\omega)=(1-\bar{P}^2)\frac{2\Gamma_c}{\omega^2+\Gamma_c^2}
\end{equation}
The autocorrelation function for the process $\Delta P'(t)$ is then~\cite{horvitz1971nuclear}
\begin{eqnarray}
R(\tau)=\Delta^2\langle P'(0)P'(\tau) \rangle &=&\Delta^2 (1-\bar{P}^2)e^{-\Gamma_c|\tau|}
\end{eqnarray}
Since the electron motion is stochastic, we are interested in the expectation value in Eq.~\eqref{eq:transition_probability} has to be taken as an ensemble average over all the possible realization of the signal $P'(t)$ from~\eqref{eq:sigma_correlation1}. Hence, we are interested in the quantity $\bigl<f(\tau)\bigr>_{P'}=\bigl< e^{-i\Delta\int_0^\tau P'(t) dt} \bigr>_{P'}$.\\
We can start from the differential equation $\dot{f}(\tau)=-i\Delta P'(\tau)f(\tau)$, which, after a single iteration, assumes the form~\cite{qubit}
\begin{eqnarray}
\nonumber
\partial_\tau f(\tau) &=& -i\Delta P'(\tau)-\Delta^2\int_0^\tau P'(\tau)P'(t)f(t) dt
\end{eqnarray}
exploiting the time reversal symmetry of $P'(t)$ we get
\begin{eqnarray}
\frac{\partial \langle f(\tau) \rangle}{\partial \tau}&=&-\int_0^\infty \theta(\tau-t) R(\tau-t) \langle f(\tau) \rangle_{P'} dt
\end{eqnarray}
in the Laplace domain, being $\Tilde{F}(s)$ the Fourier transform of $\langle f(\tau) \rangle_{P'}$, the latter equation reads $i\Tilde{F}(s)=-\Tilde{R}(s)\Tilde{F}(s)$, such that
\begin{eqnarray}
\Tilde{F}(s)=\frac{1}{\Tilde{R}(s)+s}
\end{eqnarray}
Being $\Tilde{R}(s)=\Delta^2\frac{1-\bar{P}^2}{s+\Gamma_c}$, we can easily obtain the Fourier transform of $\langle f(\tau) \rangle_{P'}$ if we use the identity for the Fourier transform of time-symmetric functions $F(\omega)=2 \text{Re}\left\{ \Tilde{F}(s=i\omega) \right\}$
\begin{eqnarray}
F(\omega)&=&\frac{2\Gamma_c  (1-\bar{P}^2)\Delta^2}{(\omega^2-(1-\bar{P}^2) \Delta^2)^2+\omega^2 \Gamma_c^2}
\end{eqnarray}
Exploiting the convolution theorem, Eq.~\eqref{eq:transition_probability} becomes
\begin{align}
W_{\ket{\alpha,\beta}\rightarrow \ket{\beta,\alpha}}&=\frac{D_{12}^2}{4}\int_{-\infty}^\infty e^{-i\Delta \bar{P} \tau-\Gamma_2 |\tau|} \langle f(\tau) \rangle_{P'} d\tau \\
&=\frac{D_{12}^2}{4} \frac{1}{2\pi}\int_{-\infty}^\infty  S_{\Gamma_2}(\omega'-\Delta\bar{P}) F(\omega')d\omega'
\label{eq:intermediate_final_trans_prob}
\end{align}
Where we have introduced the Fourier transform of $e^{-\Gamma_2|t|}$, namely
\begin{eqnarray}
S_{\Gamma_2}(\omega)&=&\frac{2\Gamma_2}{\omega^2+ \Gamma_2^2 }
\end{eqnarray}
that is, a Lorentzian shape with FWHM$=2\Gamma_2$.\\
Now, it is clear~\cite{final} that $\frac{1}{2\pi}\int_{-\infty}^\infty  S_{\Gamma_2}(\omega'-\Delta\bar{P}) F(\omega')d\omega'=2 \text{Re}\left\{ \Tilde{F}(s=i\Delta\bar{P}+\Gamma_2) \right\}$, such that~\cite{qubit}
\begin{eqnarray}
W_{\ket{\alpha,\beta}\rightarrow \ket{\beta,\alpha}}=\frac{D_{12}^2}{2} \frac{\Gamma_c(1-\bar{P}^2)\Delta^2+\Gamma_2(\bar{\Gamma}^2+\Delta^2)}{((1-2\bar{P}^2)\Delta^2+\Gamma_2\bar{\Gamma})^2+\Delta^2\bar{P}^2(\Gamma_2+\bar{\Gamma})^2}
\end{eqnarray}
where we have introduced $\bar{\Gamma}=\Gamma_{2}+\Gamma_c$.
The diffusion coefficient is then given by $D=W_{\ket{\alpha,\beta}\rightarrow\ket{\beta,\alpha}}a_{12}^2$. The final result is shown in~Fig.~\ref{fig:D_vs_Pe_1}. \\

\begin{figure}[!hbt]
\begin{subfigure}{.5\textwidth}
  \centering
  \includegraphics[width=.8\linewidth]{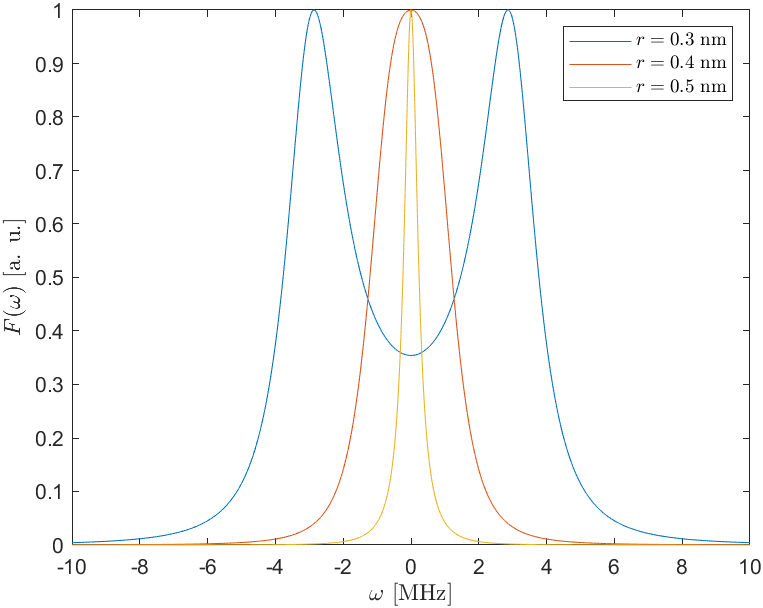}  
  \caption{}
  \label{fig:F_r}
\end{subfigure}%
\begin{subfigure}{.5\textwidth}
  \centering
  \includegraphics[width=.8\linewidth]{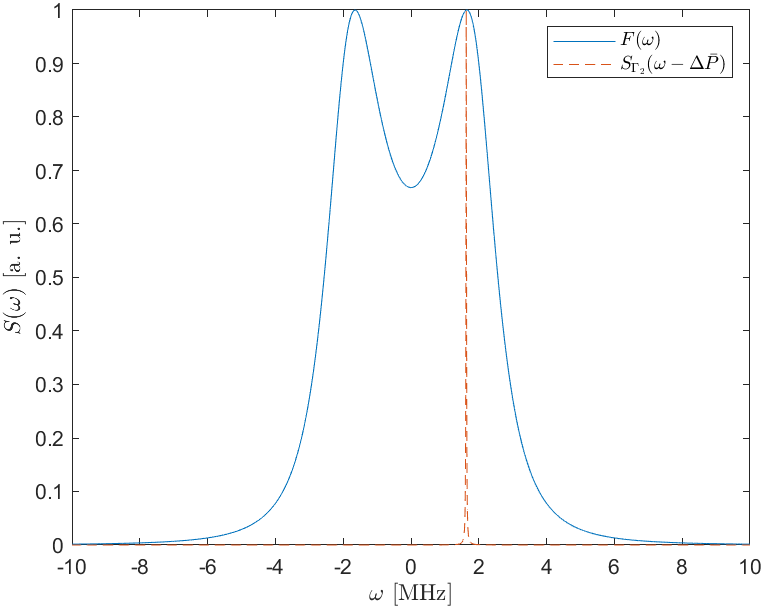}  
  \caption{}
  \label{fig:F_Gamma}
\end{subfigure}
\caption{In~(\subref{fig:F_r}) the plot of $F(\omega)$ in function of $\omega$ for different values of the electron distance.
In~(\subref{fig:F_Gamma}) the plot of $F(\omega)$ and $S_{\Gamma_2}(\omega-\Delta\bar{P})$ in function of $\omega$ at a distance from the electron of $r=\SI{0.34}{nm}$.
Both plots are obtained with the following parameters: $\bar{P}=0.6$, $\Gamma_2=\SI{9}{kHz}$, $\Gamma_c=\SI{2}{MHz}$, $a_{12}=\SI{0.66}{nm}$.}
\label{fig:fig}
\end{figure}

For $\bar{P}=0.6$, ~Fig.~\ref{fig:D_vs_Pe_1} shows that spin diffusion is particularly efficient for $r<0.5$ nm. To explain this local maximum inside the spin diffusion barrier, let us examine a specific case. In~Fig.~\ref{fig:D_vs_Pe_2}, we show the diffusion coefficient for $\bar{P}=0.6$ highlighting the position of the local maximum, at $r=\SI{0.34}{nm}$. This maximum is the result of the convolution integral of Eq.~\eqref{eq:intermediate_final_trans_prob}, which is shown in Fig.~\ref{fig:F_Gamma}. Indeed the maximum is reached whenever $F(\omega)$, that represents the spectrum of the system under the evolution of the dynamic part of $\hat{H}_{HF}(t)$, is tuned such as to have a maximum in $\Delta\bar{P}$, that is the spectrum under the static part of the hyperfine coupling.
\\

As limit cases, we can consider 
\begin{itemize}
\item The static field limit, $\Gamma_c \rightarrow \infty$ and $\bar{P}=1$, for which
\begin{eqnarray}
W_{\ket{\alpha,\beta}\rightarrow \ket{\beta,\alpha}}&=&\frac{D_{12}^2}{2} \frac{\Gamma_2}{\Gamma_2^2+\Delta^2}, \qquad \Gamma_c \rightarrow \infty
\end{eqnarray}
in agreement with the literature~\cite{suter1985spin}\cite{karabanov}
\item The high polarization limit, $\bar{P} \gg 0$. In the limit $\Gamma_2 < \Gamma_c$ the function $S_{\Gamma_2}(\omega-\Delta\bar{P})$ selects only frequencies $\omega \approx \Delta\bar{P}$, then for $\Delta > 0$ (i. e., not too far from the electron, such that $(1-P^2)\Delta^2 \ll \omega^2$) we can write
\begin{eqnarray}
F(\omega)&=&\frac{(1-\bar{P}^2)\Delta^2}{\omega^2}\frac{2\Gamma_c}{\omega^2+\Gamma_c^2}, \qquad \bar{P} \gg 0, \, \Delta>0
\end{eqnarray}
in agreement with the expression found by E. P. Horvitz~\cite{horvitz1971nuclear}.
In this regime we can take $\omega^{-2} \approx (\Delta \bar{P})^{-2}$ in the integral~\eqref{eq:intermediate_final_trans_prob} and the final result is given by
\begin{eqnarray}
W_{\ket{\alpha,\beta}\rightarrow \ket{\beta,\alpha}}=\frac{D_{12}^2}{4} \frac{1}{2\pi}\int_{-\infty}^\infty S_{\Gamma_2}(\omega'-\Delta\bar{P})F(\omega')d\omega'=\frac{D_{12}^2}{2}\frac{1-\bar{P}^2}{\bar{P}^2}\frac{\bar{\Gamma}}{\Delta^2\bar{P}^2+\bar{\Gamma}^2}, \quad \bar{P} \gg 0, \, \Delta>0
\end{eqnarray}
\end{itemize}
The results for both cases are shown in~Fig.~\ref{fig:D_vs_Pe_2}, where $D_{stat}$ corresponds to the static limit case $\Gamma_c \rightarrow 0$, $\bar{P}=1$ and $D_{dyn}$ is the result for Horvitz's approximation.


\bibliography{biblio}